\documentclass[twocolumn]{aastex631}
\shorttitle{
Cosmic Average IMF Slope Constrained by Observations
}
\shortauthors{Aoyama et al.}
\graphicspath{{./}{figures/}}
\usepackage[utf8]{inputenc}
\usepackage{amssymb, amsmath}
\date{The 27th of December 2022}
\begin{document}
\author[0000-0002-1005-4120]{Shohei Aoyama}
\affiliation{Institute for Cosmic Ray Research, The University of Tokyo, Kashiwanoha 5-1-5, Kashiwa, Chiba 277-8582, Japan}
\author[0000-0002-1049-6658]{Masami Ouchi}
\affiliation{National Astronomical Observatory of Japan, Osawa 2-21-1, Mitaka, Tokyo 181-8588, Japan}
\affiliation{Institute for Cosmic Ray Research, The University of Tokyo, Kashiwanoha 5-1-5, Kashiwa, Chiba 277-8582, Japan}
\affiliation{Kavli Institute for the Physics and Mathematics of the Universe (Kavli IPMU, WPI), The University of Tokyo, 
Kashiwanoha 5-1-5, Kashiwa, Chiba 277-8583, Japan}
\author[0000-0002-6047-430X]{Yuichi Harikane}
\affiliation{Institute for Cosmic Ray Research, The University of Tokyo, Kashiwanoha 5-1-5, Kashiwa, Chiba 277-8582, Japan}
\affiliation{Department of Physics and Astronomy, University College London, Gower Street, London WC1E 6BT, UK}
\title{
Stellar Initial Mass Function (IMF) 
Probed with Supernova Rates and Neutrino Background: \\
Cosmic Average IMF Slope is $\simeq 2-3$ Similar to the Salpeter IMF
}

\begin{abstract}
%

The stellar initial mass function (IMF) is expressed by $\phi(m) \propto m^{-\alpha}$ with the slope $\alpha$, and known as the poorly-constrained but very important function in studies of star and galaxy formation. There are no sensible observational constraints on the IMF slopes beyond Milky Way and nearby galaxies.
Here we combine two sets of observational results, 
1) cosmic densities of core-collapse supernova (CCSN) explosion rates and 
2) cosmic far UV radiation (and infrared re-radiation) densities, 
which are sensitive to massive ($\simeq 8-50 \,{\rm M}_\odot$) and moderately massive ($\simeq 2.5-7 \,{\rm M}_\odot$) stars, respectively,
and constrain the IMF slope at $m>1\,{\rm M}_\odot$
with a freedom of redshift evolution. 
Although no redshift evolution is identified beyond the uncertainties, we find that the cosmic average IMF slope at $z=0$ is 
$\alpha=1.8-3.2$ at the 95\% confidence level
that is comparable with the Salpeter IMF, $\alpha=2.35$,
which marks the first constraint on the cosmic average IMF.
We show a forecast for the Nancy Grace Roman Space Telescope supernova survey that will provide significantly strong constraints on the IMF slope with $\delta \alpha\simeq 0.5$ over $z=0-2$.
Moreover, as for an independent IMF probe instead of 1), we suggest to use diffuse supernovae neutrino background (DSNB), relic neutrinos from CCSNe. We expect that the Hyper-Kamiokande neutrino observations over 20 years will improve the constraints on the IMF slope and the redshift evolution significantly better than those obtained today, if the systematic uncertainties of DSNB production physics are reduced in the future numerical simulations.

\end{abstract}

\section{Introduction}
\color{black}
The initial mass function (IMF), $\phi(m)$, depicts the initial mass distribution of stars. 
E. E. Salpeter was the first to introduce the IMF \citep{1955ApJ...121..161S}.
He discovered that the distribution could be expressed 
by a single power-law function of stellar mass $m$ 
($\phi \propto m^{-\alpha}$) 
with a slope of $2.35$ in the solar neighborhood.
In the low-mass range of $m< 1\,{\rm M}_{\odot}$,
the following studies have claimed different IMF slopes 
(e.g. \citealt{2001MNRAS.322..231K, 2003PASP..115..763C}; 
see also \citealt{2010ARA&A..48..339B, 2013pss5.book..115K}),
however the slope in the high mass range
of $m> 1\,{\rm M}_{\odot}$ is 
similar to the Salpeter slope nearby Universe.
{ However there are some theoretical predictions which 
provide top-heavy IMFs (e.g. \citealt{2008MNRAS.385..147D}). In addition, at distant Universe, 
observational studies suggest a top-heavy IMF (e.g. \citealt{2010Natur.468..940V,2017ApJ...841...68V}).
Thus it is important to constrain the evolution of the IMF slope towards higher redshifts from observations.
}

Since stars form in molecular clouds, 
a core mass function (CMF) should be related to it.
The fragmentation process of 
the interstellar molecular clouds 
determines the CMF 
\citep{2001ApJ...559L.149I, 1997ApJ...480..681I}. 
Recently, {\it Herschel} infrared (IR) satellite identified filaments of molecular clouds 
and obtained the CMF through observations. 
In the high mass range of $m >1\,{\rm M}_{\odot}$, 
the CMF slope is almost the same as 
the Salpeter IMF slope  
 (e.g. \citealt{2010A&A...518L.102A, 2010A&A...518L.106K, 2015A&A...584A..91K}).

The IMF has been determined in nearby galaxies at a redshift of $z\simeq 0.1$ \citep{2003ApJ...593..258B} 
using optical to near-infrared (NIR) luminosity density. 
In these studies, $\alpha $ is comparable 
with the one of the Salpeter IMF, 
although the parameters in their models
degenerated with $\alpha$.
These studies are constrained in the Milky Way and low-z nearby galaxies 
since the IMF measurements need high  sensitivity.

We use observational measurements
of cosmic far ultra-violet (UV) radiation (+IR re-radiation) luminosity densities 
and cosmic core-collapse supernova (CCSN) rates 
to constrain
the cosmic average IMF in the high mass range of $m>1 \,{\rm M}_\odot$ 
over the large cosmological volume in this study (see Section 2).
The first quantity, the cosmic far UV radiation 
(+IR re-radiation) luminosity density, is sensitive to
moderately massive ($\simeq 2.5-7M_\odot$) stars and
is recognized to be an indicator of the cosmic star formation rate 
\citep{2014ARA&A..52..415M}.
The second quantity, the cosmic CCSN rate, is
sensitive to massive ($\simeq 8-50 M_\odot$) stars.
At the end of their lives, 
the massive stars with $>8\,{\rm M}_{\odot}$ 
experience CCSN explosions  
(see. \citealt{2012sse..book.....K}). 
CCSNe emit UV to optical photons 
and $\mathcal{O}(10^{58})$ neutrinos. 
Optical telescopes such as 
the Hubble space telescope detect UV and optical photons from CCSNe. 
The next generation telescope of
the Nancy Grace Roman Space Telescope (Roman telescope) will be 
extensively used to observe them.
These telescopes offer redshift measurements that help us
understand the cosmic densities of CCSN rates
as a function of redshift.
Neutrinos are good traces of CCSNe, which 
complements the CCSN rates measured by optical observations. 
Neutrinos from SN1987A were successfully identified in
Kamioka Observatory and Baksan Neutrino Observatory in February 1987
\citep{1987PhRvL..58.1490H}.
Kamioka observatory's neutrino detection with Kamiokande
imposes limits not only on the properties of the SN1987A progenitor 
but also on the many parameters beyond the standard model of elementary particles (BSMs; see. \citealt{1996slfp.book.....R}). 
Subsequently, larger neutrino observation facilities, such as Super-Kamiokande (SK), have been built, 
and various crucial physical results are reported \citep{1999NuPhS..77..123K}. 
The next generation facility, Hyper-Kamiokande (HK), is currently under construction. 
The sensitivity of HK (SK) is high enough to quantify (constrain) the numbers of
relic neutrinos from CCSNe outside the Milky Way which are 
known as the diffuse supernova neutrino background (DSNB).
The DSNB measurements will show CCSN explosions 
across the cosmic volume that can be used for 
constraining the cosmic average IMF's shape.

This paper is organized as follows. 
In Section 2, we introduce the approach of our study.
In Section 3, we present the parameter constraints derived from the existing 
observational data and forecasts of the constraints based on HK and Roman observations.
In Section 4, we discuss 
the systematic uncertainty. In Section 5,
we conclude this study.

\color{black}

\section{Method}
%

\begin{figure}
    \centering
    \includegraphics[width=85mm]{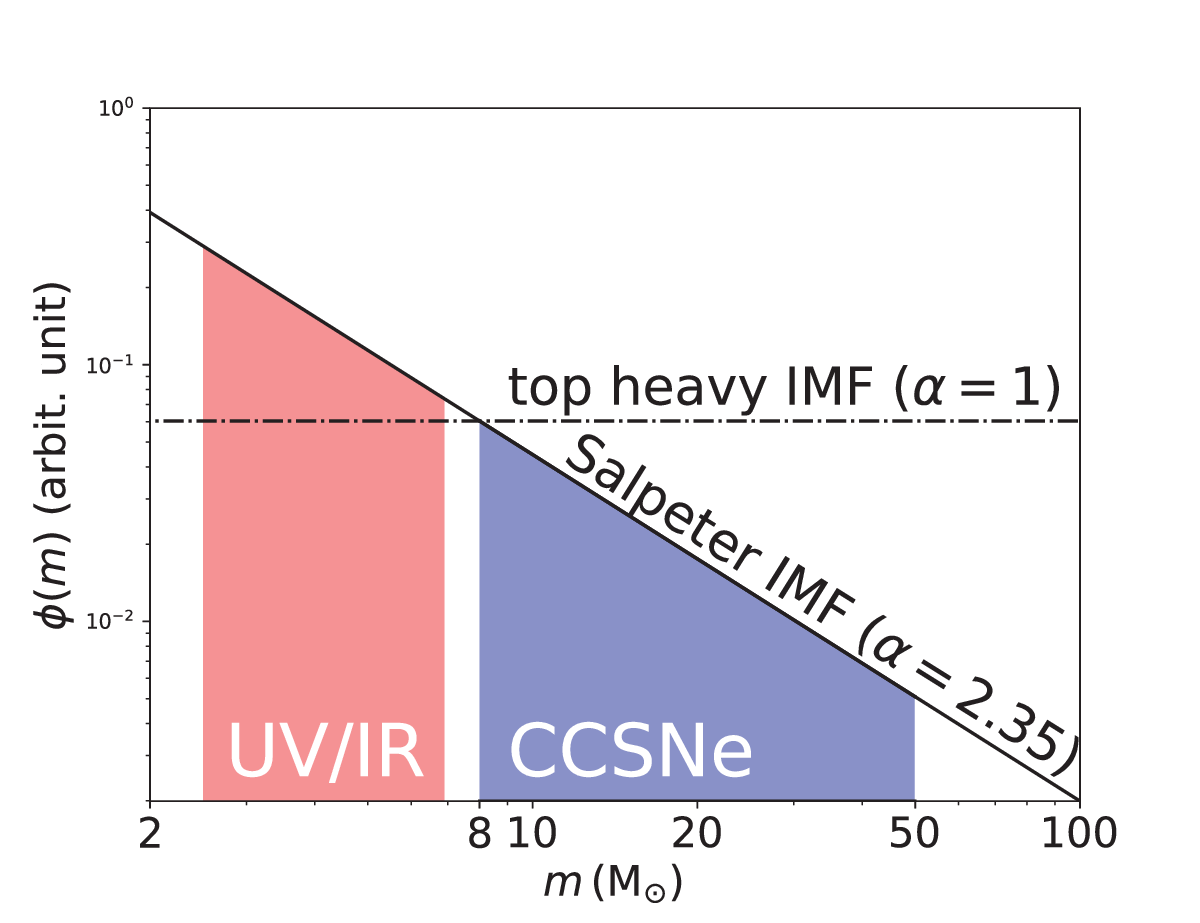}
    \caption{\textcolor{black}{
    \textcolor{black}{\bf Approximated stellar mass ranges of 
    CCSN progenitors (blue) and far UV/IR radiation contributors (red). The solid and dot-dashed lines represent the Salpter and top-heavy IMFs, respectively.
    }
    }}
    \label{fig:imf}
\end{figure}

\textcolor{black}{
We combine two sets of observational constraints 
on the IMF, 
1) cosmic CCSN explosion rates and 
2) cosmic far UV radiation (and IR re-radiation) densities, which are sensitive to massive ($\simeq 8-50 \,{\rm M}_\odot$) and moderately massive ($\simeq 2.5-7 \,{\rm M}_\odot$) stars, respectively. 
We show these mass ranges 
in Figure \ref{fig:imf}.
In this study, we constrain the IMF slope at $m>1\,{\rm M}_\odot$ with a freedom of redshift evolution. In this mass range, the IMF does not have turnovers which are found in the Milky Way (e.g. \citealt{2001MNRAS.322..231K, 2003PASP..115..763C}).
We present 1) and 2) in Sections 2.3 and 2.2, respectively.
Section 2.4 discusses 2) with DSNB for the future studies.
}

\textcolor{black}{
It should be noted that the IMF should depend on galaxy properties including metallicities, star-formation rates, and masses. In this study, 
we aim at revealing the cosmic average IMF shape at each redshift that should depend on the cosmic average metallicities, star-formation rate densities, and mass densities, using the cosmic average CCSN rates and far UV+IR densities.
It should be also noted that the time scale of the far UV and IR emission corresponds to the life-time of moderately-massive/massive stars,  \\$\sim 80\,(M\slash (7\,{\rm M}_{\odot}))^{-2.5}$ Myr \citep{2004sipp.book.....H},
where $M$ is the initial mass of 
the progenitor star.
This time scale is much shorter than the time resolution of this study (i.e. CCSN-rate data binning size; Section 4.2), $\Delta z\sim 0.1$ corresponding to $\sim 500-1000$ Myr, at $z=0-2$. The far UV and IR emission are negligibly affected by the past star formation that was taken place in the different cosmic-average galaxy properties.
}
{ For example, the results in Figure \ref{fig:my_label_0} do not change beyond the uncertainties of the IMF parameters if we vary the star formation history or stellar age.}

In the analysis of Section 2.2, sample selections of galaxies are important. In our study, we use the cosmic average far-UV/IR luminosity densities that are derived from galaxy samples with completeness corrections, integrating the luminosity function at each redshift via star-formation rate densities. The luminosity functions are complete for a given luminosity, regardless of star-formation rates, stellar ages, and masses.

\subsection{Parameterizing the IMF shape evolution}

In this study, we assume a shape of the IMF as 
a single power-law with a slope $\alpha$.
We define the minimum (maximum) mass of the IMF
as $0.5\,{\rm M}_{\odot}$ ($200\,{\rm M}_{\odot}$),
following the previous study of \citet{2003APh....18..307A}.
 

On this assumption, the functional form of the IMF becomes 

\begin{equation}
\phi(m, z ) \propto m^{-\alpha}\, (0.5\,{\rm M}_{\odot}\le m \le 200\,{\rm M}_{\odot}) 
\end{equation}

Here we include the evolution of $\alpha$ linearly depending
on a redshift.
Thus
\begin{equation}
    \alpha = \alpha_{0}+\alpha_{z}\cdot z, \label{alpha_evolution}
\end{equation}
where $\alpha_{0}$ and $\alpha_{\rm z}$ are 
free parameters. We call these parameters the IMF parameters.
In the case of the Salpeter IMF with no redshift evolution,
the parameter set is 
 $(\alpha_{0}, \alpha_{\rm z})=(2.35, 0)$.
{ Although non power-law IMFs are important (e.g. \citealt{2018MNRAS.480.4265J, 1985MNRAS.214..379L, 1976MNRAS.174..695L, 2022ApJ...931...57S,2012ApJ...760...71C, 2015ApJ...798L...4M}), these IMFs are out of the scope of the main purpose of our this study. }
 



\subsection{Star formation rate density}
A cosmic star formation rate density (SFRD) 
at a redshift $\psi (z)$ 
can be estimated 
with
far UV radiation and IR re-radiation 
densities, which are $\rho_{\rm UV}$ and 
$\rho_{\rm IR}$, respectively.
\textcolor{black}{
Because the SFRD depends on the IMF assumption,
we convert an SFRD derived with the Salpeter IMF
and the mass range of $0.1-100 M_{\odot}$ \citep{2014ARA&A..52..415M}
to those estimated with any IMFs and the mass range of
$0.5-50 M_\odot$ \citep{2009ApJ...699..486C, 2010ApJ...712..833C},
which conserves the observational quantities of $\rho_{\rm UV}$ and 
$\rho_{\rm IR}$ in the self-consistent manner.
Here
}
$\psi (z)$ is written as 
\begin{equation}
    \psi(z)=\mathcal{K}_{\rm UV}\rho_{\rm UV}+
    \mathcal{K}_{\rm IR}\rho_{\rm IR}~.
\end{equation}
The coefficients $(\mathcal{K}_{\rm UV}$, $\mathcal{K}_{\rm IR})$
are determined with the population synthesis code 
created by \cite{2009ApJ...699..486C} and \cite{2010ApJ...712..833C}.
%
Here we define an cosmic SFRD estimated with the assumption of the Salpeter
IMF $\psi^{\rm Sal}(z)$, and obtain
\begin{eqnarray}
\psi^{\rm Sal}(z)&=&
 \mathcal{K}^{\rm Sal}_{\rm UV}\rho_{\rm UV}
+\mathcal{K}^{\rm Sal}_{\rm IR}\rho_{\rm IR}\\
&=&0.015\dfrac{(1+z)^{2.7}}{1+\left[(1+z)/2.9 \right]^{5.6}}\,\,[{\rm M}_{\odot}/{\rm yr}/{\rm Mpc}^{-3}]\, ,\label{sfrd1}
\end{eqnarray}
where $\mathcal{K}^{\rm Sal}_{\rm UV}$ and 
$\mathcal{K}^{\rm Sal}_{\rm IR}$ are 
coefficients for the Salpeter IMF that are 
$3.30\times 10^{-28}\,[{\rm M}_{\odot}\,{\rm yr}^{-1}\,{\rm erg}^{-1}\,{\rm s}\,{\rm Hz}]$ and 
$3.46\times 10^{-10}\,[{\rm M}_{\odot}\,{\rm yr}^{-1}\,{\rm L}_{\odot}^{-1}]$, 
respectively. 
%

The eq. \eqref{sfrd1} is derived with the
best-fit function to many observational measurements of 
comic SFRDs
obtained by \citet{2014ARA&A..52..415M} (MD14).
%
As presented in Appendix B,
\begin{equation}
\psi(z)=
\dfrac{\mathcal{K}_{\rm UV}}{\mathcal{K}_{\rm UV}^{\rm Sal}}
\psi^{\rm Sal}(z)~,
\end{equation}
where we use the fact that the dependence of $\mathcal{K}_{\rm IR}\slash \mathcal{K}_{\rm UV}$ on $\alpha(z)$ is negligibly small, up to $\sim 3$\%.
The combination of eqs. 3 and 5 provides $\psi(z)$ as a function of redshift.

{ 
It is true that the IMF constraints from the SFR densities are affected by variables such as metallicity, typical star formation rate (SFR) and the mass of host halos at each redshift. In this paper, we do not intend to constrain the IMF redshift evolution at the fixed metallicity and SFR, but we want to discuss the IMF constraints on average at each redshift, including the effects from these parameters such as metallicity and SFR.
}




\subsection{Number fraction of SNe to all stars}

Progenitor stars of CCSNe
have a $\gtrsim 10\% $ solar metallicity and 
an initial mass within a range of 
$8\,{\rm M}_{\odot}< m <50\,{\rm M}_{\odot}$
 \citep{2003ApJ...591..288H,2003APh....18..307A}.
One can calculate the ratio of CCSNe 
to all stars, $f(\alpha)$, using the IMF with
\begin{equation}
    f(\alpha)=
    \dfrac{\int_{8\,{\rm  M}_{\odot}}
    ^{50\,{\rm M}_{\odot}}\phi(m, z)dm}
    {\int_{0.5\,{\rm  M}_{\odot}}
    ^{200\,{\rm M}_{\odot}}\phi(m, z)dm}~, \label{f_alpha}
\end{equation}
In the case of the Salpeter IMF with no redshift evolution, 
the ratio is $f(\alpha)=0.0128$. 



\subsection{Neutrino spectrum}
We use hydrodynamical simulations of CCSNe
performed by \cite{2013ApJS..205....2N} to 
estimate the neutrino spectrum 
of a CCSN\footnote{
In the simulations of \cite{2013ApJS..205....2N},
it is assumed that the mass range of the progenitor stars is 
%
$13- 50\,{\rm M}_{\odot}$, while we assume 
$8- 50\,{\rm M}_{\odot}$.
Because there are no simulation results
for
the mass range of $8-13 \,{\rm M_{\odot}}$, 
we extrapolate the 
results of 
\cite{2013ApJS..205....2N}
to
$8-13 \,{\rm M_{\odot}}$.
} 
${d^{2}\tilde{n}_{\nu}}\slash{dE_{\nu}dm }$. 
Here $E_{\nu}$ is the energy of neutrinos.
These simulations have two major parameters,  
a metallicity of the progenitor star $Z$ and 
a revival time scale of shock waves 
$t_{\rm revive}$. 
We assume $Z$ as the solar metallicity $Z_{\odot}$ 
for simplicity. We adopt $t_{\rm revive}=300\,{\rm ms}$
that is derived by observations of SN1987A \citep{2010ARNPS..60..439B}.

In the high redshift Universe, one cannot spatially 
resolve a CCSN.
\textcolor{black}{ By using} the expected spectra of the DSNB
from the simulation results of \cite{2013ApJS..205....2N}, 
we derive the IMF-weighted spectrum of neutrinos
${d\tilde{n}_{\nu}}\slash{dE_{\nu}}$
expressed as
\begin{equation}
    \dfrac{d\tilde{n}_{\nu}}{dE_{\nu}}(E_{\nu}, z)
    =\int_{8\,{\rm M}_{\odot}}
    ^{50\,{\rm M}_{\odot}}\dfrac{d^{2}\tilde{n}
    _{\nu}}{dE_{\nu}dm }\phi(m, z)dm~,
\end{equation}
\textcolor{black}{ Here we adopt the instantaneous recycling approximation, in which massive stars $>1\,{\rm M}_{\odot}$ finish their lives instantaneously \citep{1972ApJ...173...25S}}.
The spectrum of the DSNB   
$dn_{\nu}\slash {dE_{\nu}}$ is derived
as (\citealt{2003APh....18..307A,2010ARNPS..60..439B})
\begin{equation}
\dfrac{dn_{\nu}}{dE}(E_{\nu})=
\int_{0}^{+\infty}
(1+z)f(\alpha )\psi(z)
\left\{
\dfrac{d\tilde{n}_{\nu}}{dE_{\nu}}[(1+z)E_{\nu}, z]
\right\}H(z)dz\, , \label{masterEq}
\end{equation}
where $H(z)$ is the Hubble parameter at $z$. 
The term $(1+z)E_{\nu}$ corresponds to the energy of redshifted neutrinos. 
The term $f(\alpha )\psi(z)$ is the cosmic 
event rate of CCSNe 
per unit comoving volume.

The HK is sensitive to DSNBs at the energy range  
$19.7-30.0\,{\rm MeV}$ \citep{2018arXiv180504163H}.
For the HK, the neutrino flux of DSNBs $n_{\nu}^{\rm HK}$ is 
\begin{equation}
n_{\nu}^{\rm HK}=\int_{19.7\,{\rm MeV}}^{30.0\,{\rm MeV}}
\dfrac{dn_{\nu}}{dE_{\nu}}(E_{\nu})dE_{\nu}\label{spectrum}
\end{equation}
With the eq. \eqref{spectrum}, 
\cite{2018arXiv180504163H} estimate the number of DSNB neutrinos 
to be 140 
\textcolor{black}{with the 5.7 $\sigma $ 
non-zero significance level} 
under the assumption of the Salpter IMF 
with no redshift evolution and a 20-year operation of the HK.  
This number highly depends on the cosmic SFRD. 
In the \cite{2018arXiv180504163H}, 
the SFRD is significantly underestimated compared with 
recent observational results claimed in 
\cite{2014ARA&A..52..415M}, hereafter MD14. 
We expect that the HK will detect 280 DSNB neutrinos 
at the 7.4 $\sigma $ 
\textcolor{black}{non-zero significance level} 
over a 20-year operation
in the case that the cosmic SFRD is given by MD14 
under the assumptions of the Salpeter IMF with no redshift evolution . 
We estimate the neutrino flux of the 280 DSNB neutrino detections to be 
$0.33\,\pm 0.04\,{\rm cm}^{-2} {\rm s}^{-1}$,  
following the calculations of Ando et al. (2003). 
We use this neutrino flux and uncertainty to
forecast constraints on 
$(\alpha_{0}, \alpha_{\rm z})$, 
assuming this neutrino flux uncertainty independent of 
the neutrino flux.

\begin{figure}
    \centering
    \includegraphics[width=85mm]{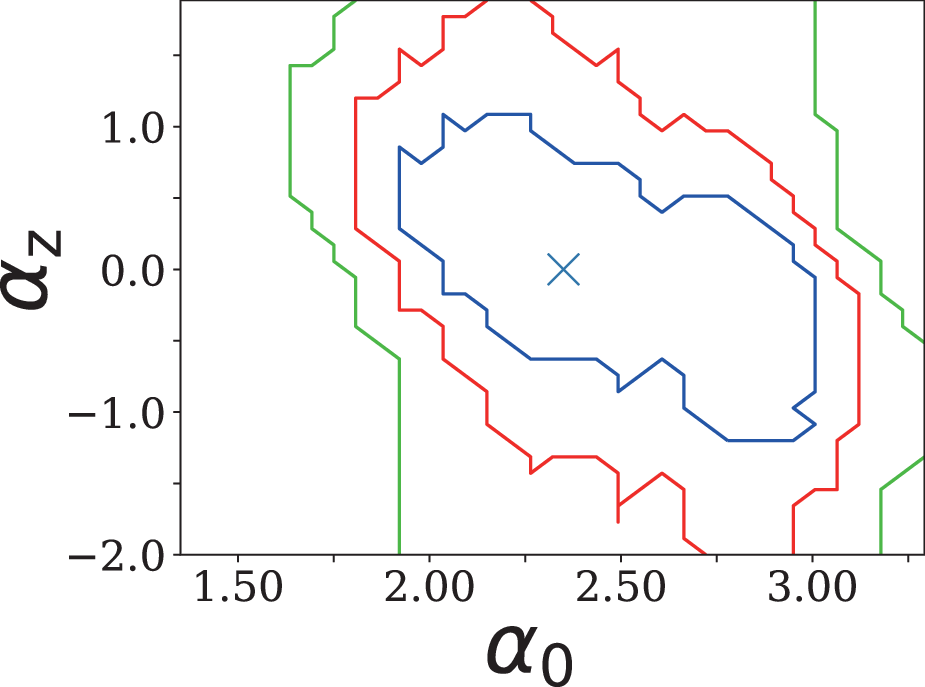}
    \caption{Constraint on a parameter set of the IMF with optical number counts of SNe up to date. The contours 
    correspond to $1, 2$ and $3\,\sigma$ CL from the inside to outside (blue, red and green lines). 
    The Salpeter IMF with no redshift evolution is represented with a cross mark "x" at the center.}
    \label{fig:my_label_0}
\end{figure}
\begin{figure}
    \centering
    \includegraphics[width=85mm]{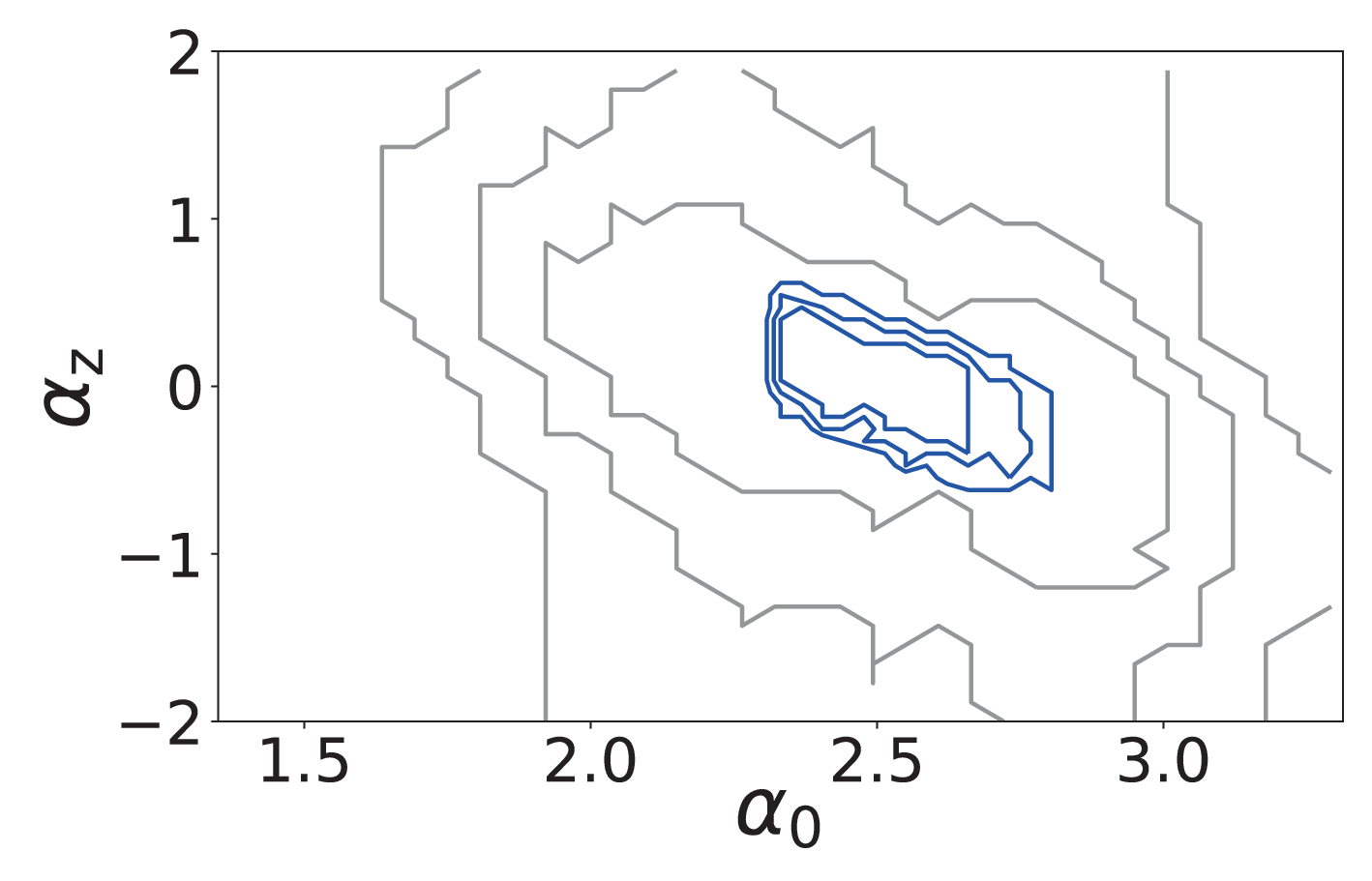}
    \caption{Forecast of constraint with 
    the Roman telescope. The contours (blue) are defined as the same as Figure \ref{fig:my_label_0}. The contours in Figure \ref{fig:my_label_0} are shown with gray lines.}
    \label{fig:my_label_1}
\end{figure}\begin{figure}
    \centering
    \includegraphics[width=85mm]{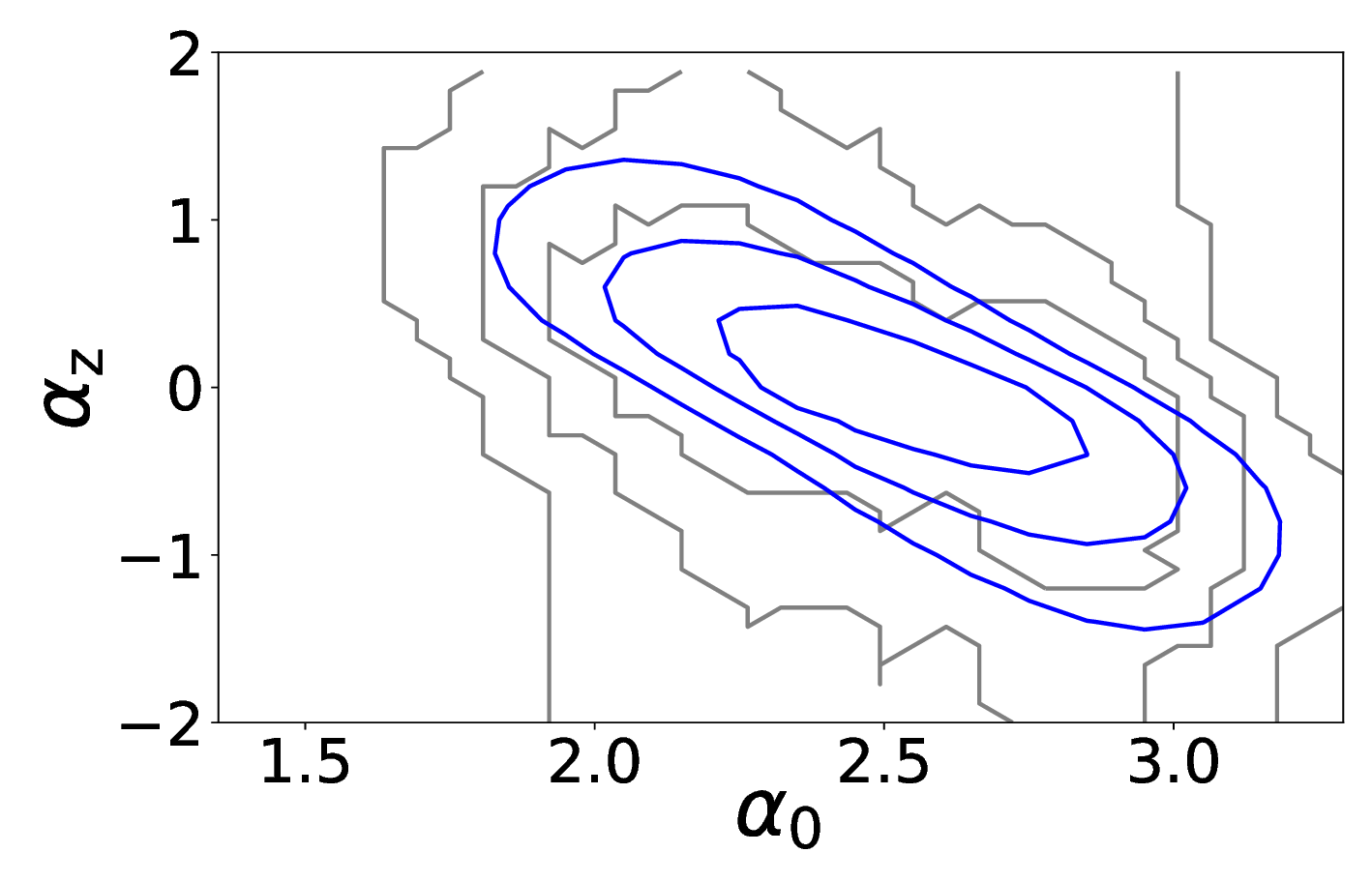}
    \caption{Forecast of constraint with 
    the HK. The contours (blue) are defined as the same as Figure \ref{fig:my_label_0}. The contours in Figure \ref{fig:my_label_0} are shown with gray lines.}
    \label{fig:my_labelNu}
\end{figure}

\section{Result}
\subsection{CCSN number counts}
\subsubsection{IMF constrained by the existing observations}
We use the CCSN number counts obtained by two observational studies of 
\cite{2012ApJ...756..111M} and \cite{2012ApJ...757...70D}, which 
directly constrain $f(\alpha )\psi (z)$ \textcolor{black}{
as a function of redshift}.
The results of these two observational studies are reliable, because the CCSNe counts are corrected for dust extinction. Moreover, more than half of the CCSNe are spectroscopically confirmed.
%
%
\textcolor{black}{By comparing the CCSNe number counts with 
our predictions, $f(\alpha)\psi(z)$, at each redshift, }
we set a constraint on the parameter space 
$(\alpha_{0},\alpha_{\rm z})$ as shown in Figure \ref{fig:my_label_0}. One can see a decreasing trend in Figure \ref{fig:my_label_0}. If $\alpha_{0}$ increases, 
$\alpha_{\rm z}$ decreases, 
\textcolor{black}{because a large number of massive stars 
are needed at high redshift to explain the observed CCSN number 
counts in the case of the higher value of $\alpha_{0}$ 
(resulting in a smaller number of massive stars at $z\simeq 0$).}
Our result is consistent with the Salpeter IMF with 
no redshift evolution albeit with the large uncertainty.
This large uncertainty originates from 
large error bars of the number counts of CCSNe
at a redshift range $z\gtrsim 0.5$ in \cite{2012ApJ...757...70D}. 
Although we only use the results of optical observations available 
up to date, 
we exclude a top (bottom) heavy IMF such as $\alpha <1.8$ 
($\alpha > 3.1$) in the local Universe at the 2 $\sigma $ CL. 
{ Our constraint is consistent 
with the previous studies such as 
\cite{2015ApJ...798L...4M} and \cite{2012ApJ...760...71C}.}

\subsubsection{IMF constrained by future observations}
The Roman Space Telescope is scheduled to be launched 
in the middle of the 2020s. 
The Roman Space Telescope has the one hundred times larger 
field of view than that of the HST. 
\cite{2018ApJ...867...23H} shows that
approximately 540 CCSNe will be identified.
We assume that the uncertainty of the number counts 
follows the Poisson statistics.
We obtain the forecast 
as shown in Figure \ref{fig:my_label_1}.
We \textcolor{black}{ can} 
narrow the $2\sigma $ error contour down to  
$2.25<\alpha_{0}<2.70$ and 
$|\alpha_{\rm z}|<0.5$. 
Here we assume the fiducial values are 
those for the Salpeter IMF with no redshift evolution.

\subsection{Neutrino fluxes measured in the future}
The HK is now under construction and 
planned to be operated from 2027.
By the 20 year operation of the HK, 
the IMF parameters are constrained 
as shown in Figure \ref{fig:my_labelNu}. 
We obtain the forecast of $2.05<\alpha_{0}<3.00$ and 
$|\alpha_{\rm z}|< 0.9$ at the 2 $\sigma $ CL.
The value of $\alpha_{\rm z}$ is constrained better than 
the one of the present SN survey results 
(Figure \ref{fig:my_label_0}).
These improvements of the parameter constraints will 
be accomplished mainly due to the increase of the detection number of
the DSNB neutrinos originated from $z\simeq 1-2$ Universe
\footnote{
The intensity peak of the DSNB falls within 
the observation window of the HK  
in the case that 
the progenitor stars reside at $z\simeq 0.5-2$.
}.

\begin{figure}
    \centering
    \includegraphics[width=85mm]{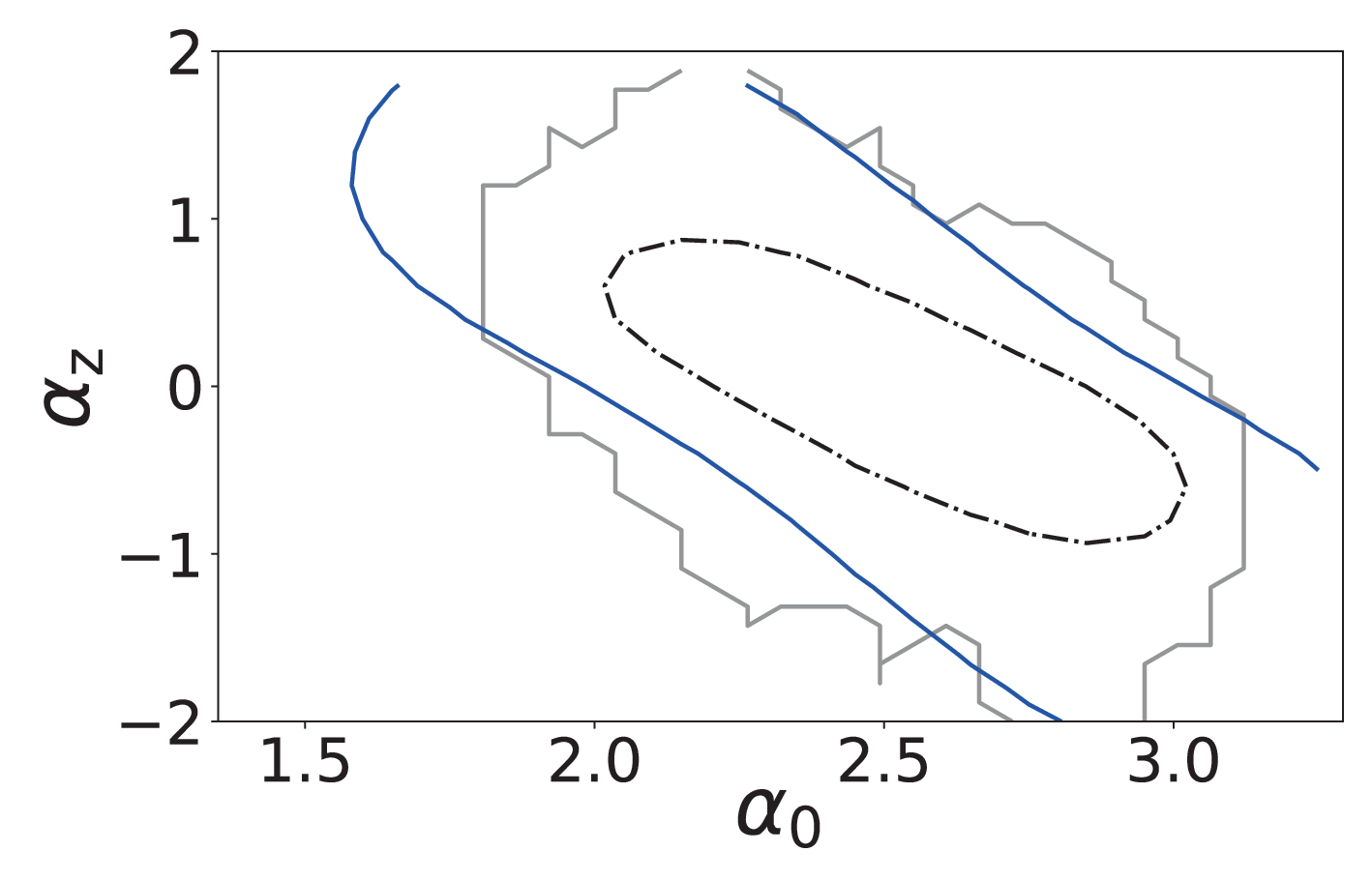}
    \caption{Uncertainty of \textcolor{black}{ the }
    parameter estimation
    \textcolor{black}{ from} the simulation uncertainty
    \textcolor{black}{ 
    (blue line)
    }. 
    All lines correspond 
    to the $2\, \sigma$ CL.
    The fiducial case is shown with black dot-dashed line,
    \textcolor{black}{ which is shown in figure 
    \ref{fig:my_labelNu}}.
    The contours in Figure \ref{fig:my_label_0} are shown with gray lines.}
    \label{fig:my_label_4}
\end{figure}

The neutrino fluxes from CCSNe can be changed by assumptions that include an approximation of gas motions around the progenitor core 
and  
a choice of the equation of state
\footnote{
Understanding of the equations of state 
is being improved by recent observations of neutron star mergers
(GW170817/GRB170817A) 
\citep{2018PhRvL.120q2703A, 2017ApJ...850L..34B} 
and very massive neutron stars 
(PSR J0740+6620) 
\citep{2020NatAs...4...72C}.
}. 
A systematic uncertainty of the neutrino flux raised by
the assumptions is estimated to be 
$\sim 50$\% 
(private communications with K. Kotake, see also 
\cite{2020MPLA...3530011M}). 
We include the systematic uncertainty into our model
calculations, and show the contours of the forecast in Figure \ref{fig:my_label_4}.
%
%
%
%
%
Figure \ref{fig:my_label_4} indicates that
one rules out the top heavy IMF 
such as $\alpha_{0} < 1.6$ with the 
$2\sigma $ CL (confidence level)
even in the case 
with 
the systematic uncertainty.
In Figure \ref{fig:my_label_4}, the forecast contours 
with the systematic uncertainty (blue line) 
is as large as the contour obtained
by this study with the optical
number counts (gray lines; Section 3.1.1).
We discuss the large uncertainty raised by the systematics in Section \ref{discussion}.2.2.

\section{Discussion}\label{discussion}
\subsection{IMF constrained by this study}
\textcolor{black}{ As shown in Section 3.1.1, } 
the cosmic IMF at $z=0$ 
falls in $1.8<\alpha <3.2$ 
with $2\,\sigma $ CL. This result excludes 
the top heavy IMF such as $\alpha = 1$ 
more than $4\, \sigma$ CL in the local Universe. 
This result suggests that the star formation mechanism in distant Universe 
is consistent with that in local Universe, which are 
claimed in \cite{2001ApJ...559L.149I} and \cite{ 2015A&A...580A..49I}.
However, the uncertainties of $\alpha_{0}$ 
and $\alpha_{\rm z}$ are large to discuss 
the IMF at high redshifts 
because the current results depend on 
only several tens CCSN observations.

\subsection{IMF constrained in the future}\label{IMF_future}
\subsubsection{CCSN number counts}

In the near future, 
the forecast of the constraint is 
$2.2<\alpha<2.7$ ($1.9<\alpha < 3.1$) at 
a redshift $z=0$ ($z=1$) . 

\subsubsection{Neutrino fluxes}
Following the discussion in Section 4.2.1,
the forecast of the HK is 
$2.25<\alpha<3.02$ $(1.35<\alpha<3.92)$
at a redshift $z=0$ ($z=1$).

Ideally, a strong constraint on the IMF parameters 
will be set as shown in Figure \ref{fig:my_labelNu}.
Note that the forecast of the IMF parameter constraints
does not include in the systematic uncertainty
raised by the theoretical model
that bridges between the CCSN explosion and the neutrino flux 
\citep{2013ApJS..205....2N}. 
If the systematic uncertainty of the theoretical model
is considered, the forecast of the IMF parameter constraints
with the future HK data is as large as the present IMF parameter constraints (Figure \ref{fig:my_label_4}). Accomplishing the parameter constraints
shown in Figure \ref{fig:my_labelNu}, one needs to reduce the systematic uncertainty
of theoretical models. One can expect that the model of the
CCSN explosion can be improved by fast-computational resources
and new computation schemes, which will narrow the systematic uncertainty.

%
%
Comparing the computational speed of the best super computers
in years of 2021 and 2001 on the basis of TOP500 ranking
\footnote{https://www.top500.org}, 
one can find that the one of year 2021
has a speed thirty six thousand times faster than 
the one of year 2001.
%
%
The computational scheme of the simulations of CCSNe 
has also been improved 
in the same period (2001-2021).
Now theorists have
realized spontaneous explosions of CCSNe, 
three dimensional simulations of CCSNe,
and the detailed neutrino transfer in the dense matter 
around the core of CCSNe, while none of them has been
conducted in 2001. 
%
Similarly, we expect to see significant improvements
in the processing speed and the calculation scheme
in the forthcoming 30 years of the HK construction
and operation. 
\textcolor{black}{
In fact, Bastian et al. (2018) \footnote{
 https://elib.dlr.de/121636/1/2018\_SuperMUC-Results -Reports.pdf}
show that it costs several million node-hours (SuperMUC/Germany) to conduct 3-D CCSN simulations, today’s state-of-the-art CCSN simulations, with full neutrino transports  (see \citealt{2020ApJ...890..127C}). If one performs 3-D neutrino transport simulations that accomplish the accuracy of 
50\% in the CCSN neutrino emission estimates with the top-ranked supercomputer today (Fugaku/Japan), 
it takes more than one years that is not realistic. Note that the Bastian et al.’s simulations took approximately 170 years with the top-ranked supercomputer 20 years ago (ASCI White/the United States), but completes in less than 2 days with Fugaku/Japan. Such a computational power evolution towards the future will allow us to realize the high-resolution magneto-hydrodynamical general relativistic neutrino-driven core-collapse supernova explosion simulations probably in $\sim 0.1$ years .
If the systematic uncertainty of the
theoretical model becomes very small,
the IMF parameter constraints of 
Figure \ref{fig:my_labelNu}
will be obtained.
}


There is a future plan that 
the second HK (HK2) will be constructed 
in Korea. 
The HK2 will be used to improve 
the detectablity of proton decays and 
the determination of the parameters of neutrinos.
The design and performance of the HK2 
are supposed to be exactly the same as those of the HK 
\citep{2018arXiv180504163H}.
If the HK2 observes DSNB neutrinos for 20 years,
the number of DSNB neutrinos detections will be doubled.
Assuming the improvements of Poisson noise by the doubled sample,
we expect that the IMF constraints will be narrowed to $2.1<\alpha_{0}<2.9$
and $|\alpha_{\rm z}|<0.7$ as shown in Figure 
\ref{fig:my_label_5}. 


\subsubsection{Implication to be obtained by the future observations}

In Section \ref{IMF_future}, we find that the IMF constraints from DSNB neutrinos ($2.25< \alpha <3.02$) are not strong as those of CCSN number counts ($2.2 < \alpha <2.7 $). 
The neutrino constraints are placed with spectra of the DSNB neutrinos integrated over redshift,
while 
CCSN number counts are obtained as a function of redshift.
Although the CCSN number counts and DSNB neutrino measurements
both provide the constraints on the IMF,
independent observational tests are needed 
to reliably understand the IMF and its evolution.

\begin{figure}
    \centering
    \includegraphics[width=85mm]{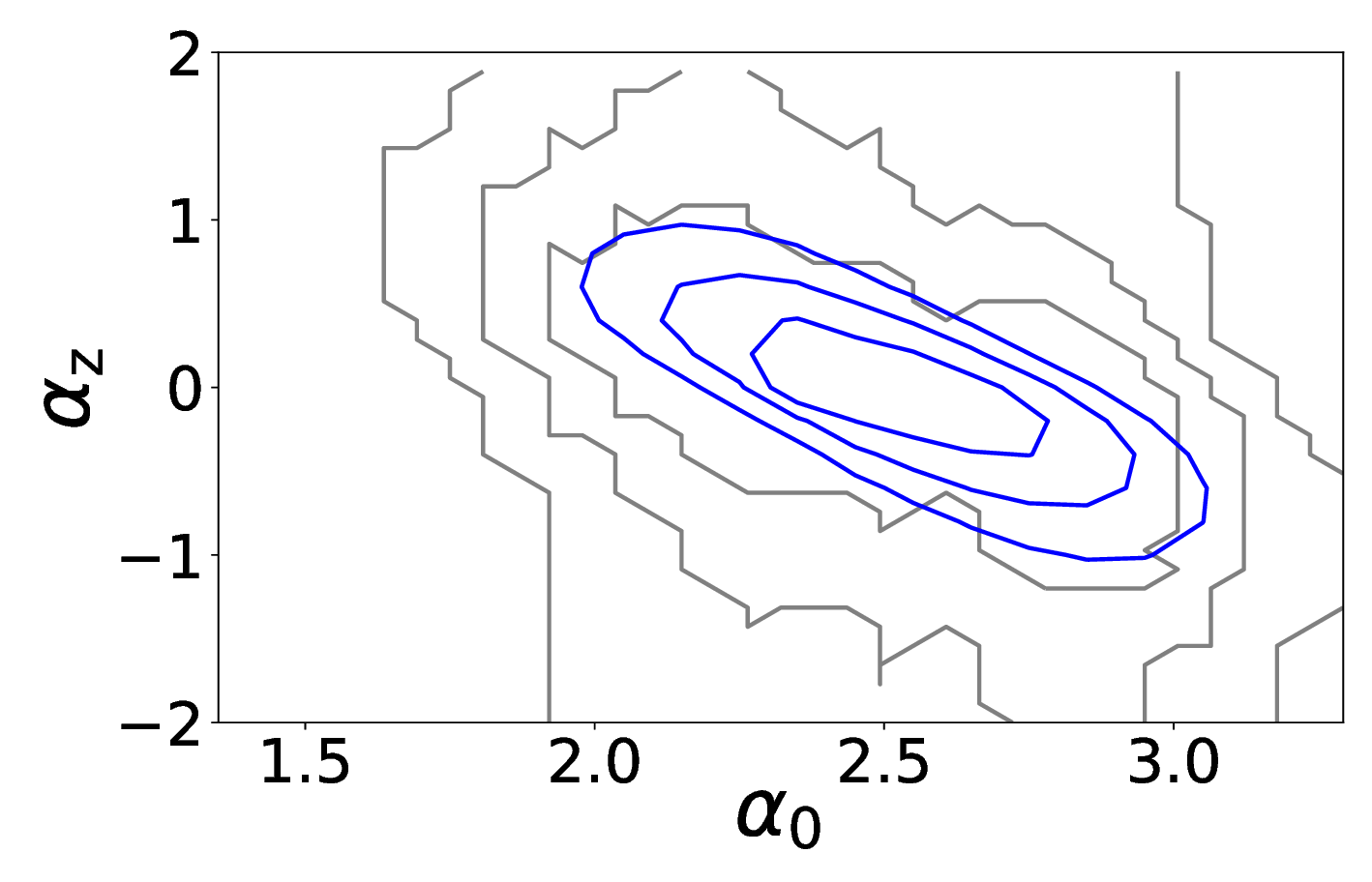}
    \caption{The forecast of constraint with the HK and the HK2. 
    The contours are defined as the same as 
    in Figure \ref{fig:my_label_0}. 
    The contours (blue) in Figure \ref{fig:my_label_0} 
    are shown with gray lines.}
    \label{fig:my_label_5}
\end{figure}

We show the forecast of constraints 
obtained with CCSN number counts of 
the Roman space telescope (Figure \ref{fig:my_label_1}) 
and 
the HK+HK2 (Figure \ref{fig:my_label_5}).
%
%
If one obtains these constraints on 
the IMF parameters in the future,
the Roman and HK+HK2 observations can test the existence of 
top-heavy IMFs of $\alpha = 1$ at $z\gtrsim 2$ such suggested in previous studies 
(e.g. \citealt{2005MNRAS.356.1191B}; \citealt{2019MNRAS.484.1852A})
\footnote{Previous studies such as \cite{2010Natur.468..940V} and \cite{2017ApJ...841...68V} 
have tested whether the IMF is top heavy (see also the theoretical work of \cite{2008MNRAS.385..147D}). Because these
studies investigate the IMF in the
low mass range of $\lesssim 1 M_\odot$ with dwarf galaxies that is different
from the top-heavy IMF arguments of massive stars ($\gtrsim 10 {\rm M}_\odot$) that we study, these previous studies have not
checked the IMF at the high mass range yet.}

.
If the HK+HK2 detects no DSNB neutrinos,
the HK+HK2 will rule out the Salpeter IMF 
with no redshift evolution. 

\section{Conclusion}

%

We suggest a new method to constrain average IMF shapes of galaxies.
We include the linear evolution of 
\textcolor{black}{the IMF slope,} 
$\alpha $, 
by redshift.
The IMF parameters can be constrained by 
the CCSN number counts and DSNB neutrino numbers. 
With the currently available observational data of 
the CCSN number counts, we obtain $1.8 < \alpha_{0} < 3.1$ 
at the $2\,\sigma $ CL (Figure \ref{fig:my_label_0}). 
The CCSN number counts suggest that the IMF 
in the local Universe is consistent with
the Salpeter IMF with no redshift evolution.
Moreover, top-heavy IMFs in the local Universe 
are ruled out by the CCSN number counts.
%
%
However, redshift evolutions of the IMF shape are not identified
with the current observational data due to 
the small sample of the CCSNe
%
in the previous studies.

In the future, Roman space telescope will identify
CCSNe ten times larger than 
the number of currently available CCSNe.
The large number of CCSNe will determine 
the IMF parameters with high accuracy (Figure \ref{fig:my_label_1}).
The HK and HK+HK2 can also place a constraint on 
the IMF parameters with the number of DSNB neutrinos
detected over the 20-year operation (Figure \ref{fig:my_labelNu}).
%
%
This constraint is independent from the one of the CCSN number counts. 
If there are no systematic errors more than those 
suggested in \cite{2012PhRvD..85e2007B}
%
we expect to place the constraint on the IMF parameters 
over the 20-year operation of the HK 
(Figures \ref{fig:my_labelNu} and \ref{fig:my_label_5}).
With these constraints, one can test 
the top heavy IMF at high redshift such 
\textcolor{black}{suggested} by 
\cite{2005MNRAS.356.1191B} and \cite{2019MNRAS.484.1852A} .

\section{acknowledgements}
\textcolor{black}{
We acknowledge an anonymous referee for 
providing useful comments and discussions.
Numerical computations were carried out on analysis servers 
and Cray XC50 at the Center for Computational Astrophysics (CfCA), National Astronomical Observatory of Japan, 
Yukawa-21 at the Yukawa Institute Computer Facility in Kyoto University. SA acknowledges the CfCA
for providing the computing resources of analysis servers and Cray XC50.
This work is supported by the World Premier International
Research Center Initiative (WPI Initiative), MEXT, Japan, as
well as KAKENHI Grant-in-Aid for Scientific Research (A)
(20H00180 and 21H04467) through the Japan
Society for the Promotion of Science (JSPS).}
YH acknowledges support from the JSPS KAKENHI grant No. 19J01222 and 21K13953.
This work is supported by the joint research program of the Institute for Cosmic Ray Research (ICRR), the University of Tokyo.
\newpage 
\section{Appendix}
By changing the IMF, the estimated $\psi(z)$ from 
UV/IR flux is changed. By using the fact that 
a coefficient 
$\mathcal{K}_{\rm IR}\slash \mathcal{K}_{\rm UV}$ 
does not depend on $\alpha$ with a few percent accuracy,
we can obtain $\psi(z)$ from $\psi_{\rm Sal}(z)$ as follows.

\begin{eqnarray}
\psi(z)&=&
\mathcal{K}_{\rm UV}
\rho_{\rm UV}
+\mathcal{K}_{\rm IR}
\rho_{\rm IR}\\
&= &\mathcal{K}_{\rm UV} \rho_{\rm UV}
+(\mathcal{K}_{\rm IR}\slash \mathcal{K}_{\rm UV})
\mathcal{K}_{\rm UV} \rho_{\rm IR}\\
&\simeq &\mathcal{K}_{\rm UV} \rho_{\rm UV}
+(\mathcal{K}_{\rm IR}^{\rm Sal}\slash \mathcal{K}_{\rm UV}^{\rm Sal})
\mathcal{K}_{\rm UV} \rho_{\rm IR}\\
&= &\mathcal{K}_{\rm UV}/\mathcal{K}_{\rm UV}^{\rm Sal}
\left( 
\mathcal{K}_{\rm UV}^{\rm Sal} \rho_{\rm UV}
+\mathcal{K}_{\rm IR}^{\rm Sal} \rho_{\rm IR}
\right)\\
&=&\mathcal{K}_{\rm UV}/\mathcal{K}_{\rm UV}^{\rm Sal}
\psi^{\rm Sal}(z)\, .
\end{eqnarray}

\bibliography{imf2021b}
\end{document}